\documentclass[a4paper]{jpconf}

\usepackage{amssymb}
\usepackage{amsmath}
\usepackage{graphicx}
\usepackage{dcolumn}
\usepackage{epsfig}
\usepackage[dvips]{color}
\usepackage{bm}
\usepackage{fancybox}
\usepackage{mathrsfs}
\usepackage{booktabs}
\usepackage{framed}
\definecolor{shadecolor}{gray}{0.90}

\begin{document}
\begin{flushright}
MPP-2011-144
\end{flushright}
\title{Light neutralino in the MSSM: An update with the latest LHC results}

\author{Lorenzo Calibbi$^1$, Toshihiko Ota$^1$ 
and Yasutaka Takanishi$^{2,}$\footnote[3]{Present address: Max-Planck-Institut f\"{u}r Kernphysik,
Saupfercheckweg 1, 69117 Heidelberg, Germany}}
\address{$^1$Max-Planck-Institut f\"{u}r Physik (Werner-Heisenberg-Institut)\\
F\"{o}hringer Ring 6, 80805 M\"{u}nchen, Germany}
\address{$^2$Physik-Department, Technische Universit\"{a}t M\"{u}nchen,
85748, Garching, Germany}
\ead{calibbi@mppmu.mpg.de, toshi@mppmu.mpg.de, yasutaka@mpi-hd.mpg.de}

%
%

\begin{abstract}
We discuss the scenario of light neutralino dark matter 
in the minimal supersymmetric standard model, which is 
motivated by the results of some of the direct detection 
experiments --- DAMA/LIBRA, CoGeNT, and CRESST.
We update our previous analysis with the latest results 
of the LHC. 
We show that new LHC constraints disfavour
the parameter region that can reproduce the 
results of DAMA/LIBRA and CoGeNT.
\end{abstract}

\section{Introduction}
One of the most dramatic progresses in our understanding of nature
in the last decade is the establishment of the dark sector of 
the universe, which was brought by precision cosmological 
observations.
However, even basic properties --- mass and interactions --- 
of Dark Matter (DM) have not been fully understood yet.
Numerous experimental efforts to detect DM have been made, 
and some of the direct detection experiments claim a positive 
signal that might be originated by DM.
The DAMA/LIBRA and CoGeNT collaborations show an annual modulation 
of signal events, which can be interpreted as the change of 
the relative velocity of the detectors against the DM halo.
In this conference, the long-awaited result of the CRESST 
collaboration was presented,
which indicates a significant excess of 
events~\cite{talk-CRESST}.
The results of these experiments 
suggest a light DM field ($M_{\chi} \sim 10$ GeV) with 
a relatively large scattering 
cross-section with nucleons ($\sigma \sim 10^{-41}$ cm$^{2}$);
see e.g.,~\cite{talk-Arina}.
In this talk, we discuss the scenario of light neutralino 
DM in the minimal supersymmetric standard model (MSSM).
Since the lightest supersymmetric particle is protected by R-parity
and thus is stable, the lightest neutralino is a good candidate 
for DM.
This possibility has been extensively discussed in the prior 
studies~(cf. \cite{talk-Munoz,talk-Fornengo} and references therein).
In our recent paper~\cite{Calibbi:2011ug}, we inspected 
the compatibility in the MSSM parameter space between 
the light neutralino DM scenario and particle physics 
constraints such as 
$B$ and $K$ meson decays, 
neutralino production rate,
and 
Higgs boson searches.
Here, we update our analysis with the latest results of the LHC.
We will show that the light neutralino DM scenario
that is consistent with the results of DAMA/LIBRA and CoGeNT
comes into disfavour with the latest LHC bounds, especially 
constraints from the neutral Higgs boson search.

\section{Light neutralino in the MSSM}
What does a light neutralino with mass of around 10 GeV 
require of the MSSM parameters?
The mass of the neutralino is mainly controlled by three parameters,
the Bino mass $M_{1}$, Wino mass $M_{2}$, and higgsino mass $\mu$. 
Since two of them, $M_{2}$ and $\mu$, are related to chargino masses,
they must be larger than $\sim 100$ GeV to satisfy their LEP mass bounds.
Therefore, 
the only possibility is to take $M_{1} \sim 10$ GeV 
to obtain the light neutralino. 
Consequently, the lightest neutralino in this scenario 
becomes mostly Bino.
The model that we deal with is described by the following 
nine parameters:
\begin{align}
\tan \beta, \quad
M_{1}, \quad
M_{2}, \quad
M_{3}, \quad
a_{0}, \quad
\mu, \quad
m_{A}, \quad
m_{\tilde{q}},\quad
m_{\tilde{\ell}},
\label{eq:parameters}
\end{align}
where $\tan\beta$ is defined as the ratio of the two Higgs VEV's, 
$M_3$ is the gluino mass, $a_{0}$ is the universal coefficient for
tri-linear scalar couplings, $m_A$ is the CP-odd Higgs mass,
and $m_{\tilde{q}}$ and $m_{\tilde{\ell}}$ are the common soft SUSY
breaking masses for squarks and sleptons, respectively.
We scan them to search for the parameter region
in which the light neutralino DM scenario is consistent
with the particle physics observations\footnote[1]{%
We use the public codes {\sf SuSpect}~\cite{Djouadi:2002ze}, 
{\sf micrOMEGAs}~\cite{Belanger:2006is}, 
and {\sf SuperIso}~\cite{Mahmoudi:2008tp} in 
our numerical studies.}.

We assume that the neutralino was thermally produced in the early
universe. Since such a light particle is overproduced in the thermal process, 
an efficient annihilation process is necessary to reproduce the correct 
relic density.
There are two main annihilation processes for light neutralino DM:
Sfermion mediation and CP-odd Higgs mediation.
It is known that the parameter space in which sfermion mediation 
becomes the main annihilation process cannot be consistent with 
the results suggested by DAMA/LIBRA, CoGeNT 
and CRESST~\cite{Dreiner:2009ic} (see also Ref.~\cite{Vasquez:2011yq,Cumberbatch:2011jp}).
Therefore, following Ref.~\cite{Fornengo:2010mk}, 
we adopt the Higgs mediation process
as the main annihilation process.
In order to enhance the Higgs-mediated annihilation,
the following three conditions are required of the model parameters: 
(i) light $m_{A}$, (ii) small $\mu$ to have a significant amount of 
higgsino component in the lightest neutralino, and (iii) large $\tan \beta$ 
to enhance the main annihilation process to a $b\bar{b}$ pair.
This parameter choice enhances not only the annihilation process
but also many flavour physics processes. 
In addition, it magnifies the signal of neutral Higgs bosons 
decaying to lepton pairs at hadron colliders. 
Therefore, one must carefully inspect the compatibility in the
parameter choice between the cosmological requirements and
the particle physics constraints. 

\begin{figure}[t]
\unitlength=1cm
\begin{picture}(16,6)
\put(0,6){\includegraphics[height=8cm,angle=-90]{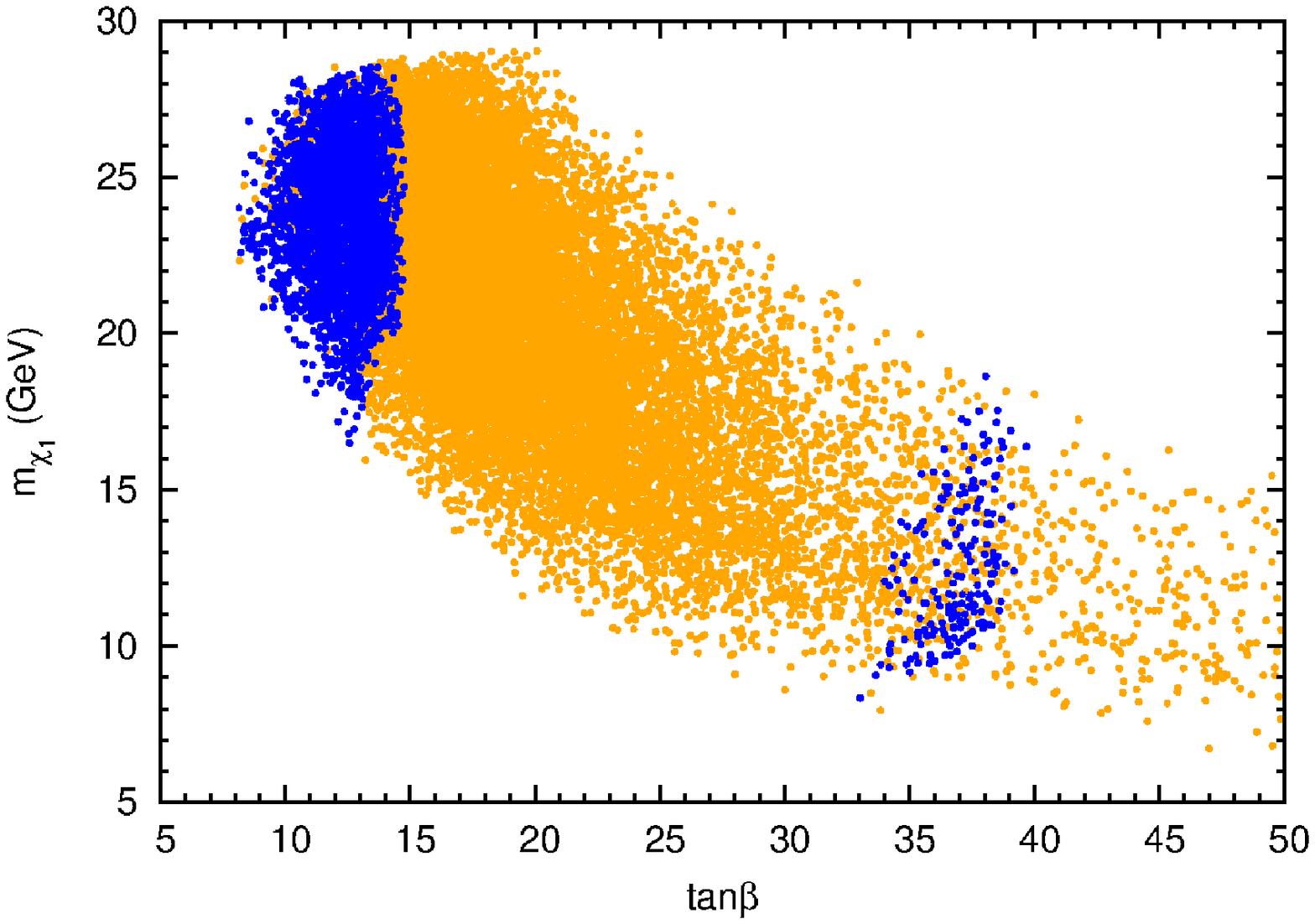}}
\put(8,6){\includegraphics[height=8cm,angle=-90]{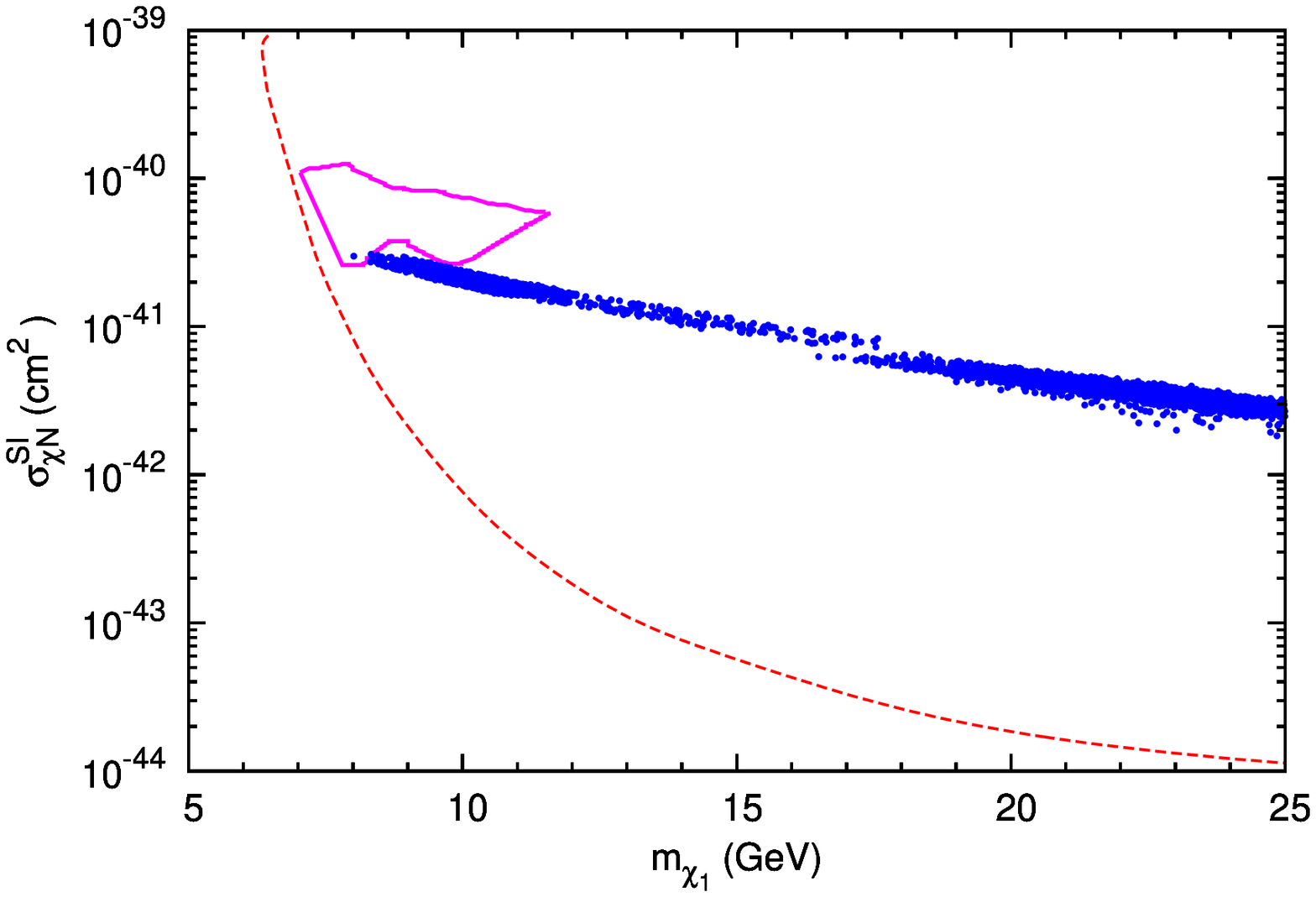}}
\put(14,1.7){\footnotesize Xenon100}
\put(10.4,4.9){\footnotesize CoGeNT}
\end{picture}
\caption{Left: 
Allowed parameter region on the plane of the lightest neutralino 
mass and $\tan \beta$. After taking into account all the constraints,
two separated regions are left (blue points):
(i) A high $\tan \beta$ strip and (ii) A low $\tan \beta$ region.
To obtain a light neutralino with $M_{\tilde{\chi}^{0}_{1}}< 15$ GeV,
a large value of $\tan \beta$ is necessary.
Right:
Neutralino-nucleon scattering cross-section as a function of 
mass of the lightest neutralino.
The prediction of the high $\tan\beta$ strip 
roughly coincides with the results of DAMA/LIBRA and CoGeNT.
The best-fit region of CRESST is located around 
$(M_{\tilde{\chi}^{0}_{1}}, \sigma_{\rm SI})\simeq$(20 GeV, 
$10^{-42}$ cm$^{2}$)~\cite{talk-CRESST}. 
}
\label{Fig1}
\end{figure}

We employ the following flavour physics processes
and categorize them into two groups by their parameter dependence: 
\begin{itemize}
\item {\bf Group I:} 
$B \rightarrow \tau \nu$, $B \rightarrow D \tau \nu$,
$D_{s} \rightarrow \tau \nu$, and $K \rightarrow \mu \nu$.
They essentially depend only on Higgs parameters,
$m_{H^{\pm}}$ and $\tan \beta$.

\item {\bf Group II:}
$b\rightarrow s\gamma$ and $B_{s} \rightarrow \mu^{+} \mu^{-}$.
These processes depend not only on the Higgs parameters but also 
on the SUSY parameters such as $m_{\tilde{q}}$ and $a_{0}$.
\end{itemize}
Group I observables complemetarily constrain the parameter space of
$m_{H^{\pm}}$ and $\tan \beta$, which are regardless of the detail 
of the SUSY parameters~\cite{Antonelli:2008jg}. 
The key observables are $B\rightarrow \tau \nu$ and 
$K \rightarrow \mu \nu$. We adopt the following values  
\begin{align}
0.52 < R_{B\tau \nu} < 2.61
\quad
\text{and}
\quad
0.985< R_{\ell 23}(K \rightarrow \mu \nu) 
< 1.013,
\end{align}
as the allowed range of these observables.
$B\rightarrow \tau \nu$ excludes intermediate values of $\tan \beta$. 
On the other hand, $K \rightarrow \mu \nu$ disfavours 
larger values of $\tan \beta$.
Therefore, the allowed parameter region with middle-to-high values 
of $\tan \beta$ is sharply narrowed by combination of these 
two constraints.
After a numerical scan of the parameters Eq.~\eqref{eq:parameters},
we found that the allowed parameter regions form
two separated clusters: 
(i) A narrow strip at high $\tan \beta$ ($32\lesssim\tan \beta\lesssim38$)
and (ii) A low $\tan \beta$ region ($7\lesssim \tan \beta \lesssim 15$).
The mass of the lightest neutralino is inversely proportional to
$\tan \beta$, because a lighter neutralino requires a more efficient 
annihilation process that is enhanced by a large value of 
$\tan \beta$.
As shown in the left plot of Fig.~\ref{Fig1},
a neutralino with mass lighter than 15 GeV
is viable only on the high $\tan \beta$ strip.

Next, we discuss
the neutralino-nucleon scattering cross-section.
DAMA/LIBRA and CoGeNT suggest relatively large values for the cross-section.
In both parameter regions (i) and (ii),   
the neutralino-nucleon scattering process is mediated by Higgs bosons,
and its parameter dependence is similar to 
that of the neutralino annihilation process. 
Consequently, the direct detection cross-section 
becomes large at the light neutralino region on which 
a large annihilation cross-section is required.
As shown in the right plot in Fig.~\ref{Fig1},
the parameter choice of the high $\tan \beta$ strip 
approximately reproduces the results of DAMA/LIBRA and CoGeNT.

\begin{figure}[t]
\unitlength=1cm
\begin{picture}(16,6)
\put(0,-0){\includegraphics[width=8cm]{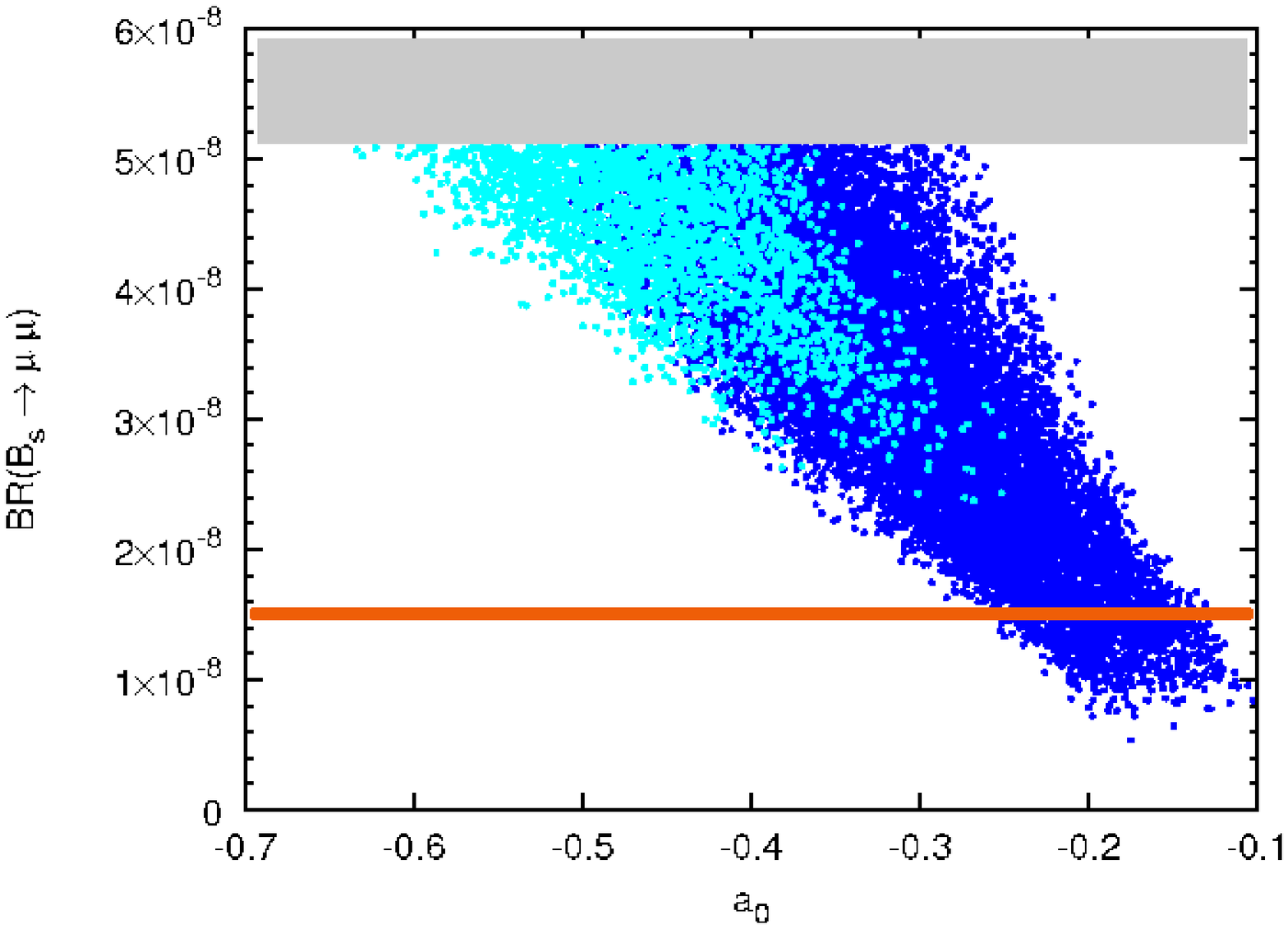}}
\put(8,0){\includegraphics[width=8cm]{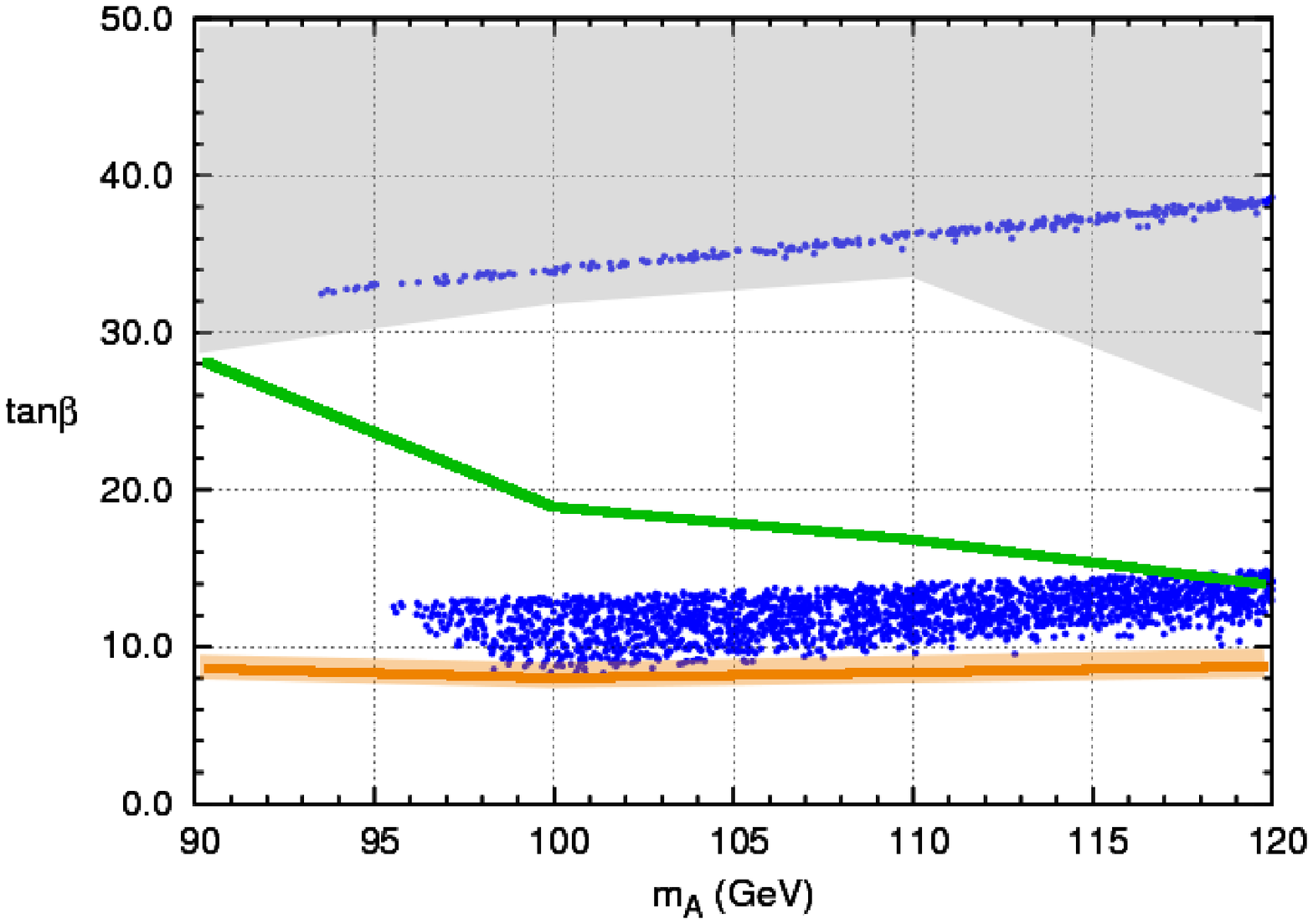}}
\put(1.8,5){\footnotesize Excluded by D$\emptyset$}
\put(1.8,2.2){\footnotesize Excluded by LHCb 0.3 fb$^{-1}$}
\put(9.5,5.2){\footnotesize Excluded by D$\emptyset$ 7.3 fb$^{-1}$}
\put(13.2,2.9){\footnotesize Excluded by}
\put(13.,2.6){\footnotesize ATLAS 1.06 fb$^{-1}$}
\put(13.6,1.3){\footnotesize CMS 1.6 fb$^{-1}$}
\end{picture}
\caption{Left: Branching ratio of $B_{s} \rightarrow \mu^{+}\mu^{-}$ 
as a function of $a_{0}$ with the points on the high $\tan \beta$ strip.
Light blue points satisfy the LEP bounds on Higgs boson masses, which 
are excluded by the latest LHCb result (orange line).
Right: Allowed points overlayed  with the constraints to the Higgs 
parameters $m_{A}$ and $\tan \beta$ from the neutral Higgs boson search 
at hadron colliders.
The latest results of both CMS and ATLAS (also D$\emptyset$) 
exclude the high $\tan \beta$ strip at 95\% CL.}
\label{Fig2}
\end{figure}

Since the high $\tan \beta$ strip is standing on the edge of the 
experimental constraints with delicate tune of 
the model parameters Eq.~\eqref{eq:parameters},
it might be easily excluded by an improvement of some 
of the constraints.
Here, we update our previous analysis~\cite{Calibbi:2011ug} with the latest LHC results.
The bound to the branching ratio of $B_{s} \rightarrow \mu^{+} \mu^{-}$ 
is improved by LHCb with an integrated luminosity 
of 0.3 fb$^{-1}$~\cite{Bettler:2011rp}, which is constrained to be smaller 
than $1.5 \times 10^{-8}$ at 95 \% CL.
As shown in the left plot in Fig.~\ref{Fig2},
this new bound excludes all the points in the high $\tan \beta$ region,
which satisfy the Higgs mass bounds at LEP.
Another crucial constraint is brought from the search for 
the neutral Higgs bosons decaying to lepton pairs. 
The latest results of CMS~\cite{CMS} and ATLAS~\cite{ATLAS} 
with an integrated luminosity of around 1 fb$^{-1}$ 
shave a large area of the high $\tan \beta$ region, as
shown in the right plot in Fig.~\ref{Fig2}.
The successful run of the LHC even allows us 
to access the low $\tan \beta$ region.

\section{Conclusions}
We have discussed the scenario of light neutralino dark matter
in the MSSM, which is motivated by the direct detection experiments 
DAMA/LIBRA, CoGeNT, and CRESST. 
In our previous study, we showed that 
a narrow parameter region with high values of $\tan \beta$ 
was consistent with the results of DAMA/LIBRA and CoGeNT, and 
it could satisfy the cosmological requirements 
and the particle physics constraints.
In this talk, we have updated our numerical analysis with 
the latest LHC results. 
It turns out that 
the high $\tan\beta$ region comes into disfavour with 
the new bound to $B_{s} \rightarrow \mu^{+} \mu^{-}$ reported by LHCb.
In addition,
the neutral Higgs boson search at CMS and ATLAS
brings a critical constraint to this scenario, 
which excludes a large part of the parameter region 
with small $m_{A}$ and large $\tan \beta$.
There is still discussion on the background estimation for the 
relevant process~\cite{Baglio:2011xz}, 
and thus we should wait for the consensus.
It is expected that the next official release from LHC
will finally examine the viability of this scenario.

\section*{References}

\providecommand{\newblock}{}

\end{document}